\documentclass[a4paper,twocolumn]{article}
\usepackage[affil-it]{authblk}
\usepackage{graphicx}
\usepackage{fullpage}
\usepackage{listings}
\title{\textbf{PNPCoin:\\ Distributed Computing on Bitcoin infrastructure}}
\author{Martin Kol\'a\v{r}%
\thanks{Electronic address: \texttt{kolarmartin@fit.vutbr.cz}}}
\affil{Brno University of Technology}
\date{}
\begin{document}

\maketitle

\abstract{
Research and applications in Machine Learning are limited by computational resources, while 1\% of the world's electricity goes into calculating 34 billion billion SHA-256 hashes per second\cite{narayanan2018blockchain}, four orders of magnitude more than the 200 petaflop power of the world’s most powerful supercomputer. The work presented here describes how a simple soft fork on Bitcoin can adapt these incomparable resources to a global distributed computer. By creating an infrastructure and ledger fully compatible with blockchain technology, the hashes can be replaced with stochastic optimizations such as Deep Net training, inverse problems such as GANs, and arbitrary NP computations.
}

\section{Introduction}

The existence of a powerful global computing infrastructure is paramount in the deployment of Artificial Intelligence. There exists a global computing system, the blockchain~\cite{nakamoto2008bitcoin}. Based on anonymous SHA-256 hash computation, it is a unified ledger of verified transactions. In addition to being secure, Bitcoins are distributed for calculating the hashes necessary to keep the system running, which creates sufficient incentive for millions of people to participate in the global computation. In fact, they perform more computations than the combined power of the 500 most powerful supercomputers by several orders of magnitude. Although the hashes may be run on purpose-made hardware which is not Turing-complete, the blockchain computes an estimated 1000 billion billion floating point operations per second\footnote{Hashes and FLOPS cannot be compared rigorously, we consider 20 FLOPS per hash, but this can be 20000 on a modern CPU}, or 1000000 petaflops, compared to 200 petaflops of Summit, which cost 300 million dollars to build. If the general-purpose computers used to compute the blockchain were available for research, the scientific community could harness the computational power of 50'000 such supercomputers.

This proposal outlines the limitations of existing Volunteer Computing projects, explains how the Bitcoin technology may be adapted to perform useful computations, and lists several problems that can currently only be solved with this new shared global computer. 

Never has a Volunteer Computing infrastructure of such magnitude been created, and this work presents a way to harness this power. The creation of another project of success comparable to Bitcoin is unlikely, but there are Bitcoin Improvement Proposals\cite{BIP} in place to allow for updates of the existing framework. This work proposes updating the proof-of-work algorithm used in Bitcoin, currently the SHA-256 hash function, with a flexible jash function allowing useful computations for the benefit of global science. By allowing SHA-256 hashes as well as new general computations, this improvement can be created without jeopardizing the existing Bitcoin structure, only requiring agreement on a soft fork to the consensus rules. 

The proposed update, called PNPCoin, enables solving certain types of otherwise intractable computational problems to the blockchain, and receive results within minutes. Several of these problems are critical to AI and Deep Learning, such as finding the next optimum in hyperdimensional stochastic gradient descent, computing the inverse of a nonlinear deep network, and finding the appropriate input to a Generator to fit a Discriminator in GAN applications. Among other applications is brute-force theorem proving, such as running Sledgehammer\cite{paulson2010three} on randomly generated theorems, a critical step in superhuman problem solving.

\section{Volunteer Computing}

Citizen science of the digital era gave rise to volunteer computing, the sharing of computational resources for research. The earliest projects, such as the Great Internet Marsene Prime Search (1996), distributed.net (1997), and SETI@home (1999) shared custom programs to be run on client machines, using a custom server to communicate data and results. However, these widely successful projects were limited to perform a single task.

Recognizing the need to perform a wide range of computations, the Berkeley Open Infrastructure for Network Computing created a generalized platform for distributed applications, enabling research in mathematics, linguistics, medicine, molecular biology, climatology, environmental science, and astrophysics. BOINC includes a verified credit system which distributes credits for performed computations.

Six years after the introduction of blockchain technology, Gridcoin\cite{halford2014gridcoin} developed a modified proof-of-stake timestamping system called proof-of-research to reward participants for computational work completed on BOINC. However, by relying on the BOINC infrastructure, submitting new computations still requires the creation of client code for every platform, and the set-up and maintenance of a BOINC server. There are only 35 such projects, none supporting general-purpose computations.

Bitcoin dwarfs these projects, both in terms of the amount of computation and the decentralized original engineering effort.

\section{PNPCoin}

The PNPCoin Soft Fork for Bitcoin enables general NP-complexity computations using volunteer computing, performed as they are submitted for a turnaround of minutes. The \textit{hash} of Bitcoin is replaced by the \textit{jash} function, which is an arbitrary piece of code which much satisfy the following requirements:

\begin{enumerate}
	\item It compiles with the current gcc\footnote{GNU Compiler Collection}
	\item It is deterministic across runs, architectures, and compilations
	\item It accepts a single binary argument, \textbf{arg} of length $n$ bits
	\item It returns a single $m$-bit string \textbf{res} of $0$s and $1$s for a given input
	\item It cannot contain while loops or recursion, and every loop is limited to run $s$ times. This guarantees a $O(n^c)$ time complexity. It will be shown that all programs can be converted into this form
\end{enumerate}

Every \textit{jash} is accompanied by a meta file, and an optional data bundle. The data must be available online, and its checksum included in the meta. The meta file defines how data is acquired (direct download/peer to peer filesharing) and extracted/decompressed, and optionally which portion of data to retrieve for a given $arg$. The \textit{jash} does not communicate with the internet, but it may include references to third-party libraries, such as CUDA\textregistered, DAKOTA, or ROOT.

\begin{figure*}[h!]
\centering
\includegraphics[width=0.55\textwidth]{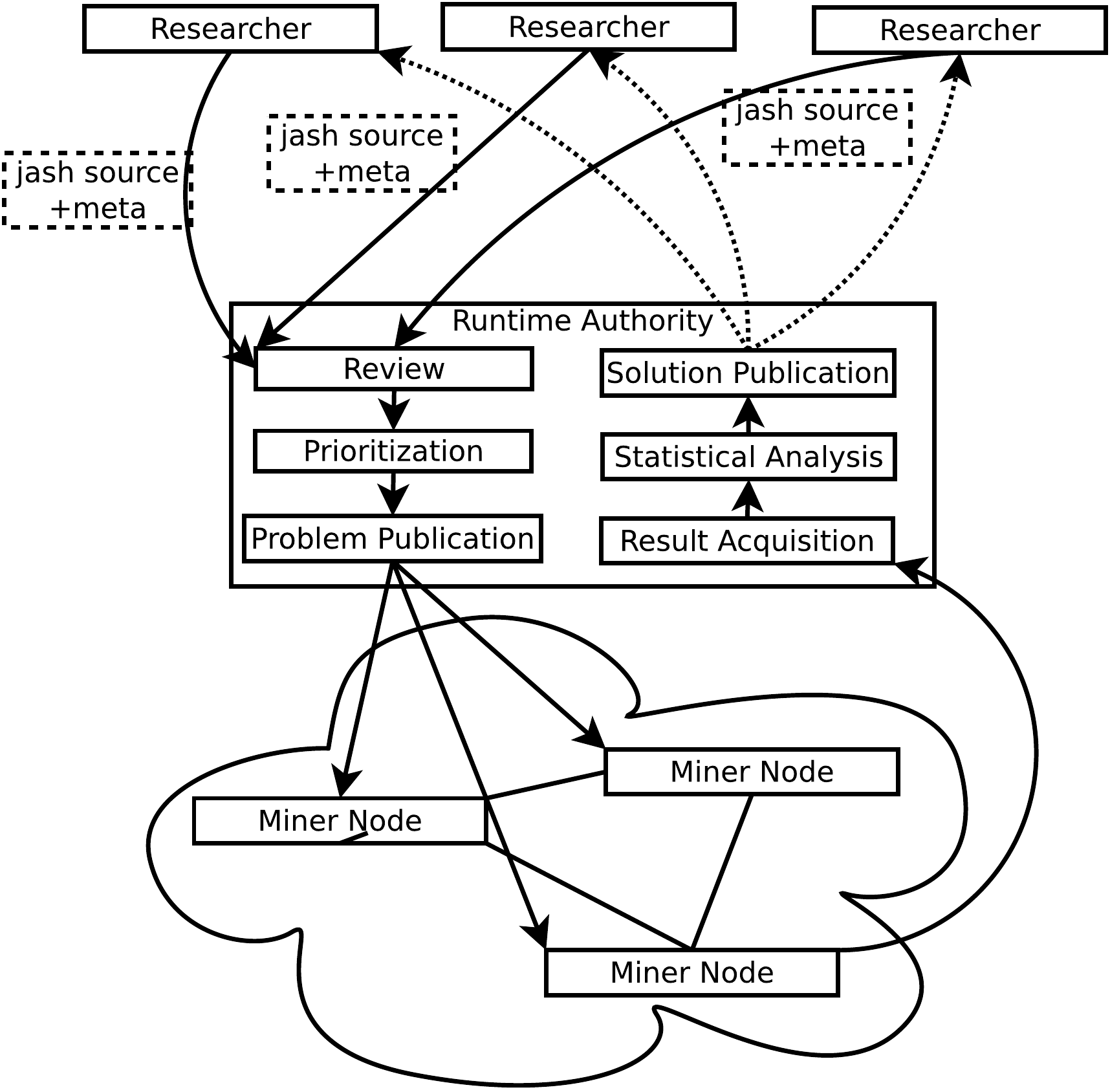}
\caption{The role of the Runtime Authority is to review code submitted by researchers, publish jash functions to be used at a given block, and aggregate results. It does not intervene in the ledger or blockchain.}
\label{structure}
\end{figure*}

As outlined in Figure~\ref{structure}, researchers submit their \textit{jash} functions to the Runtime Authority (RA), for review, prioritization, and publication. Published functions are computed for every valid input argument \textbf{arg} by miners, who then submit concatenated plain results with hashed results for every valid input argument back to the RA. The RA collects the outputs, and returns them to each researcher.

\subsection{Blockchain Integration}

The PNPCoin proposal aims to replace the SHA-256 hashing function at the core of Blockchain technology with useful computations. Instead of calculating the hash function of a randomized input, a one-way jash function is distributed at every block, and calculated instead of the hash. As in the original, results are shared by nodes communicating the hash of the blockchain, transactions are signed by new owners private keys, and timestamps are performed by distributed voting by proof-of-work.

The jash replaces the hash only in the proof-of-work step. In order to achieve greater granularity than powers of two as Bitcoin does, the jash meta can contain an upper bound on the \textbf{arg} (eg.: $n=16$ and max(\textbf{arg})=47'000).

\subsection{Bounded Complexity}
\label{bounded}

Nested recursion among functions can be flattened into a single recursive function, and all recursive functions can be converted into unbounded loops. Loops which terminate after an unpredictable number of steps are replaced with for loops with a fixed upper bound, and a break statement is added for early termination. Figures~\ref{before} and \ref{after} demonstrate this concept for the Collatz conjecture:

\begin{figure}
\begin{lstlisting}[language=C++]
b = 37
while(b!=1){
	if(b%2==0)
		b=b/2
	else
		b=3*b+1
}
\end{lstlisting}
\caption{Unbounded complexity code}
\label{before}
\end{figure}

\begin{figure}
\begin{lstlisting}[language=C++]
b = 37
for (i = 1; i<=s; i++){
	if(i==s)
		output 111
		exit
	if(b==1)
		break
	else if(b%2==0)
		b=b/2
	else
		b=3*b+1
	
}
\end{lstlisting}
\caption{Conversion to bounded complexity}
\label{after}
\end{figure}

\subsection{Runtime Authority}

While Runtime Authority resources are available via peer to peer file hosting shared by research institutions and volunteers, the content is aggregated according to rules outlined here, a review process, and additional rules created by a committee. This structure is similar to IANA, which manages Internet resources~\cite{iana}.

Each submitted \textit{jash} is validated by checking whether it compiles, and estimating mean runtime and deviation by performing runs on random inputs. The functions are prioritized according to upper bound complexity (calculated at compile time), the upper bound time complexity, data size $d$, average and deviation runtime estimates, importance ($0$ to $1$), and a veto to prevent malicious use. All but the last two criteria are fully automated, allowing fast turnaround.

There are two modes of execution for \textit{jash} functions: \textbf{full} and \textbf{optimal}. Optimal execution accepts the lowest \textbf{res}, that is, the result with most leading zeros. Full execution returns the output of every valid input to the RA, and the reward is distributed evenly across all first submissions of results. The RA is also responsible for providing a reference node implementation.

\subsection{Back-compatibility}

In order to satisfy the requirements for a Soft Fork, the proposed system is compatible with the current SHA-256 hashes. For all historic blocks, the RA will publish jash functions containing the SHA-256 hashes with fixed input, and empty meta files. In the future event that candidates are unavailable for computation, these Classic problems (SHA-256 hashes) will be published.

The RA aggregates all outputs of all sequences.
\section{Use case - Cellular Docking}

A researcher registered with the Runtime Authority wishes to test $N_p$ peptide chains for molecular docking on $N_r$ cell receptors. They begin by writing a C++ matcher for a single peptide chain - cell receptor pair, which runs in the order of milliseconds. Then, they map the space of pairs to a binary sequence

\begin{equation}
b = (n_r\bmod{N_r} + n_p*N_r)_2
\end{equation}

The space of $b$ is $n = N_r * N_p$, and the length of $b$ is at most $log_2(n)$. The code accepts an argument $b$ of $log_2(n)$ characters, left padded with zeros for constant length.

Next, they define a binary output of size $m$, in this case $m$=2, with outcomes $01$, $00$ and $10$ only, for \textit{binds}, \textit{does not bind}, and \textit{did not terminate}. This is because of the upper bound on the number of steps in every for loop, where the code may be forced to gracefully terminate prematurely.

The code is then converted not to contain while statements and recursive calls, as per section~\ref{bounded}. All peptide chains and receptor molecules are saved in the data file, made available online, along with a meta file containing its checksum.

This code is submitted to the Runtime Authority for review and publication. Here, it is compiled, tested for runtime, and the upper bound is calculated. Once all tests are passed, and the code is selected for the following block, the source and data is circulated on peer-to-peer filesharing under a unique ID. Nodes download the code, execute it, and return the outcomes to a peer-to-peer fileshare. For \textbf{optimal} execution, the first lowest solution is accepted as included in the blockchain timestamp. For \textbf{full} execution, the input and output are hased with SHA-256, and the longest leading zeros are rewarded, in addition to a smaller reward to every first submitter.

Once all results are collected, the next block begins. The input and output of the optimal solution are saved on the fileshare for \textbf{optimal} mode, and all outputs in the case of \textbf{full} mode. 

\section{Conclusion}

The results are publicly available, contributing to transparency and reproducibility. Furthermore, the community of Bitcoin miners is very creative, and has often discovered clever ways to solve hard problems. If a general less computationally intensive way to solve the given problems is discovered, it will be of even greater benefit.

The PNPCoin Soft Fork has two limitations: long processes requiring a large amount of internal memory, such as the Lucas-Lehmer primality test cannot be performed on this architecture, because they are inherently unparallelizable. Only problems which can be performed in a single step (brute-force search) or multiple optimization steps (hyperparameter tuning) are applicable. Secondly, jash functions are computed on a one-per-block basis, putting an inconvenient limitation on the runtime on each node. Resolving this would require some form of ledger allowing steps of varying length, with some computations taking a second, and others a month.

However, other otherwise intractable problem types can be easily molded into this architecture. This enables the solution of large tests over discrete hyperparameters, distributed training, hyperspace mapping, and generally pushing the solvability of NP-problems by several orders of magnitude. AI applications which require building a multi-million dollar supercomputer for will benefit 50 thousand times more from pushing Citizen Science to the Bitcoin. Thanks to the prevalence of Bitcoin, scaling AI to a global supercomputer is now realistic, and may provide a crucial step toward General Artificial Intelligence.

\bibliography{bibliography.bib}{}

\begin{thebibliography}{1}

\bibitem{halford2014gridcoin}
Rob Halford.
\newblock Gridcoin: Crypto-currency using berkeley open infrastructure network
  computing grid as a proof of work, 2014.

\bibitem{iana}
ICANN.
\newblock The iana functions: An introduction to the internet assigned numbers
  authority (iana) functions.
\newblock 2015.

\bibitem{BIP}
Eric Lombrozo.
\newblock Bip classification.
\newblock {\em https://github.com/bitcoin/bips/blob/master/bip-0123.mediawiki},
  2015.

\bibitem{nakamoto2008bitcoin}
Satoshi Nakamoto.
\newblock Bitcoin: A peer-to-peer electronic cash system.
\newblock 2008.

\bibitem{narayanan2018blockchain}
Arvind Narayanan.
\newblock Hearing on energy efficiency of blockchain and similar technologies.
\newblock {\em United States Senate, Committee on Energy and Natural
  Resources}, 2018.

\bibitem{paulson2010three}
Lawrence~C Paulson and Jasmin~Christian Blanchette.
\newblock Three years of experience with sledgehammer, a practical link between
  automatic and interactive theorem provers.
\newblock In {\em PAAR@ IJCAR}, pages 1--10, 2010.

\end{thebibliography}
\bibliographystyle{plain}

\end{document}